# Microwave photonic crystal with tailor-made negative refractive index

P. Vodo, P. V. Parimi, W. T. Lu, and S. Sridhar
*Electronic Materials Research Institute and Physics Department, Northeastern University, 360 Huntington Avenue, Boston, MA 02115*

R. Wing
*Air Force Research Laboratories, Hanscom AFB, Massachusetts 01731.*

Negative refraction and left-handed electromagnetism in a metallic photonic crystal (PC) wedge are demonstrated in free space for both transverse magnetic and electric mode propagation. The experimental results are in excellent agreement with numerical calculations based on the band structure with no fit parameters used in modeling. The results demonstrate precision control on the design and fabrication of the PC to achieve tailor-made refractive indices between -0.6 and +1.

Left-handed electromagnetism and negative refraction have recently been observed in the microwave frequency range in composite metamaterials [1] made of split ring resonators and wire strips, and in photonic crystals (PCs).[2,3] Negative refraction in Left-handed metamaterials (LHM), when used effectively, opens the door for new approaches to a variety of applications from microwave to optical frequencies. An interesting application of negative refraction is superlensing [4] effect by a flat lens with no curved surfaces that can potentially overcome the diffraction limit imposed by conventional lenses. Indeed a flat lens without optical axis has been fabricated [5] recently using a photonic crystal structure. However, for device applications it is important to have control over material parameters to be able to design and predict material properties.

Both negative refraction and left-handed electromagnetism have been demonstrated in PCs using a parallel plate waveguide.[6,7] The two-dimensional (2D) parallel plate structures used in all previous experiments to demonstrate negative refraction confine the left-handed material and lead to spurious edge effects as observed in Ref. 6. Close to the surface of refraction, the emerging wavefronts interfere, and a clear negatively refracted beam is difficult to observe. Also, the coupling of the photonic crystal modes with the incident wave is different for transverse magnetic (TM) and electric (TE) mode propagations and is not well achieved in parallel plates. In addition, bandwidth is a crucial element for device applications in a wide frequency range. It is therefore essential to investigate left-handed electromagnetism and negative refraction in a PC suspended in free space for both TM and TE mode propagations.

In this letter we report negative refraction for both TM ($\vec{E} \parallel$ to the rod axis) and TE ($\vec{E} \perp$ to the rod axis) mode propagation, in a metallic PC prism suspended in free space. Results show that a PC can exhibit negative refraction with tailor-made negative indices in a large frequency range. The propagation in different bands of the PC can be tuned with frequency to obtain either negative or positive refraction. Thus the present tailor-made PC can be utilized for a variety of applications.

The microwave PC consists of an array of cylindrical copper tubes of height 60 cm and outer radius 0.63 cm arranged on a triangular lattice. The ratio of the radius $r$ to the lattice constant $a$ was $r/a = 0.2$. Refraction experiments were performed in an anechoic chamber of dimensions 5 x 8 x 4 m$^3$ to prevent reflections from the walls. A square X-band horn placed at 3 m from the PC acts as a plane wave source (Fig. 1). Placing a piece of microwave absorber with a 6x6in.$^2$ aperture in front of the PC narrows the incident beam. On the far side another square horn attached to a goniometer, swings around in two-degree steps to receive the emerging beam. Refraction is considered positive (negative) if the emerging signal is received to the right (left) of the normal to the surface of refraction of the PC. Measurements were carried out with the incident wave vector $\vec{k}_i$ along $\Gamma \rightarrow M (0,1)$ direction of the first Brillouin zone of the PC and in both TM and TE modes. The angle of incidence $\theta = 60°$ for $\Gamma \rightarrow M$ is chosen in order to minimize surface periodicity along the surface of refraction, thus eliminating higher order Bragg waves.

Fig 2(a) shows a plot of the transmitted intensity measured at different angles and incident frequencies for the TM mode propagation. As can be seen from the figure between 6 and 7.1 GHz the signal emerges on the positive side of the normal to the surface corresponding to positive refraction. No transmission is observed between 7.1 and 8.3

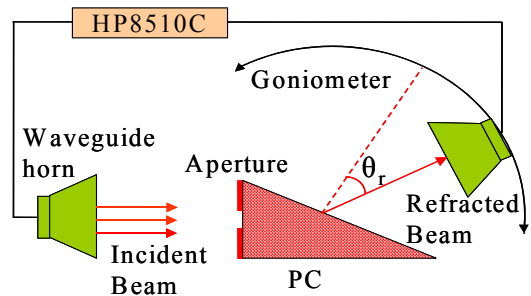

**Fig. 1** Microwave free space refraction experiment set up in an anechoic chamber. Negative or positive refraction is determined from the direction of the emerging signal with the normal to the surface of refraction.

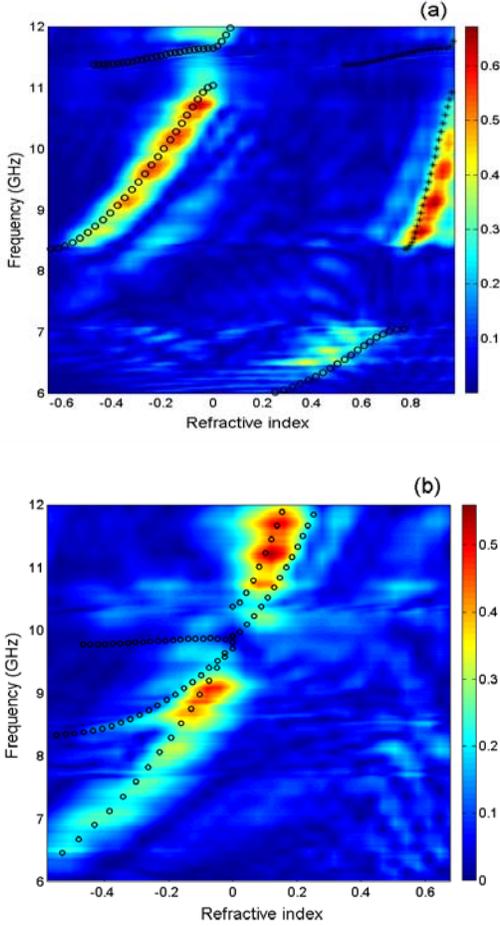

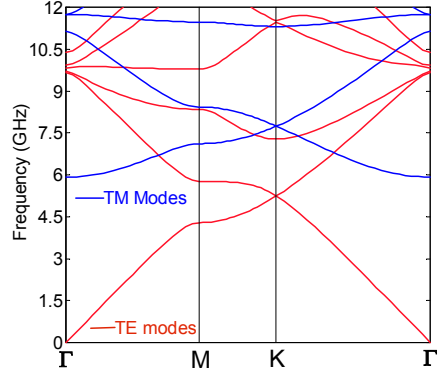

**Fig. 2** (a) Plot of refracted wave intensity measured at various angles, for TM mode propagation. (b) Similar plot for TE mode. Note the negative and positive refraction observed in different bands. Circles: Theoretically calculated refractive indices corresponding to 0th order Bragg wave and stars 1st order Bragg wave, both of which match strikingly with the experimental results without any fitting parameters.

GHz. Above 8.3 GHz up to 11 GHz two signals are observed both on the positive and negative sides of the normal. The negatively refracted signal is strongest around 10.7 GHz and positively refracted signal around 8.6 GHz. Although both positively and negatively refracted signals are observed, with the increase in frequency, positive signal gets weaker while negative signal gets stronger. The experimental refractive index $n$ is obtained from applying Snell's law $n = \sin(\theta_r)/\sin(\theta_i)$ to each emerging beam. The validity of Snell's law has been established earlier in metallic PCs. [6]

We have also carried out measurements of refraction for TE mode propagation. The results for this mode are shown in Fig. 2(b). Here, negative refraction is observed between 6.4 – 9.8 GHz and positive refraction between 9.8 -12 GHz. It is important to note that negative refraction is possible for both TM and TE modes; such a freedom in the choice of modes provides a crucial advantage of using the metallic PC over the split ring wire array metamaterial.

**Fig. 3** Band structures of the triangular metallic PC for both TM (dashed lines) and TE (solid lines) propagation modes are presented.

The physical principle behind the present results can be understood from the band structure of the metallic PC. We have calculated the band structure of the triangular lattice PC employing a standard plane wave expansion method using 2500 plane waves. The 2D band structures of both TM and TE are shown in Fig. 3. For a plane wave with incident wave vector $\vec{k}_i$ and frequency $\omega$ incident normally on an air-PC interface, the wave vector $\vec{k}_f$ inside the PC is parallel or anti-parallel to $\vec{k}_i$ as determined by the band structure (Fig. 4). For a general case the phase and group velocities in a medium are $\vec{v}_p = (c/|n_p|)\hat{k}_f$ with $\hat{k}_f = \vec{k}_f/|\vec{k}_f|$ and $\vec{v}_g = \nabla_{\vec{k}}\omega$. Note that the direction of the group velocity $\vec{v}_g$ in an infinite PC coincides with that of the energy flow. Conservation of the $\vec{k}_f$ component along the surface of refraction would result in positive or negative refraction, depending on whether $\vec{k}_f$ is parallel or anti-parallel to group velocity.

The emerging beam can be written as
$\Psi_t = a_0 e^{i\vec{k}_{t0}\cdot\vec{r}} + a_1 e^{i\vec{k}_{t1}\cdot\vec{r}}$ where $\vec{k}_{t0}$ and $\vec{k}_{t1}$ represent the refracted wave vectors corresponding to the zero and first - order Bragg wave vectors of the field inside the PC. Let the component of $\vec{k}_t$ along the normal to the surface of refraction of the prism be $k_{t\perp}$. One has
$k_{t0\perp} = \sqrt{\omega^2/c^2 - k_f^2 \sin^2\theta}$ and
$k_{t1\perp} = \sqrt{\omega^2/c^2 - (2\pi/a - k_f \sin\theta)^2}$ where $\theta = \pi/3$ and $a$ is the lattice periodicity. The parallel component is $k_{t\|} = k_f \sin\theta$. The refracted angle can be obtained from
$\theta_r = \tan^{-1}(k_{t\|}/k_{t\perp})$ for each beam.

From the band structure and the equi-frequency surfaces (EFS) for TM mode of propagation, negative refraction is predicted for the second and third band regions, with positive refraction in the first band. In the first band between 6-7.1GHz the EFS move outward with increasing frequency, so that $\vec{v}_g \cdot \vec{k}_f > 0$. In the second band between 8.3-11 GHz, the EFS move inward with increasing frequency, consistent with $n_p < 0$, corresponding to $\vec{v}_g \cdot \vec{k}_f < 0$ ($\vec{v}_g$ antiparallel to $\vec{k}_f$). The band gap is in the frequency range 7.1-8.3 GHz between the first and second pass bands and from 11-11.2 GHz between the second and third bands.

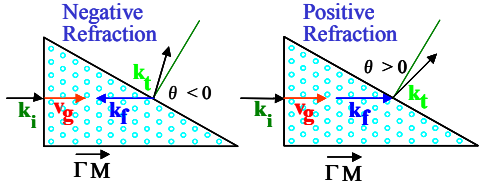

**Fig. 4** Directions of incident transmitted and refracted wave vectors, and group velocity inside the PC. Positive or negative refraction can be observed depending on whether $\vec{v}_g \cdot \vec{k}_f > 0$ or $\vec{v}_g \cdot \vec{k}_f < 0$.

An effective refractive index can be defined $n_p = \text{sgn}(\vec{v}_g \cdot \vec{k}_f) c |\vec{k}_f| / \omega$ and calculated from the band structure. The sign of $n_p$ is determined from the behavior of the EFS. In Figs. 2(a) and 2(b) the refractive indices for various bands determined from the theory are plotted. The close match between the theoretical predictions and experimental data is striking. All the predicted features including band gap, negative and positive refraction are observed in the experiment. The degeneracy observed in case of TE mode is due to multiple bands for a single frequency, which results in multiple propagation *k* vectors. It is an interesting observation that different intensities are associated with different propagation vectors for TE mode. For this mode and for $\Gamma \rightarrow M$ propagation from Fig. 3, it can be seen that the higher the slope of the curve the more intense the beam. A particular feature of interest is the bandwidth for negative refraction and lefthanded electromagnetism. From Figs. 2(a) and 2(b) it can be deduced that the bandwidth for TE mode is 42 % and for TM mode 27%. In comparison, a relatively weakly modulated dielectric PC has a bandwidth estimated to be 6% (Ref. 3) which is very narrow and the experimentally obtained bandwidth for metamaterials [1,8] to date is only 10% (Ref. 8). The present bandwidths for both TE and TM modes are higher than that in metamaterial. Bandwidth puts stringent restrictions on the tunability and functional range of the devices based on the LHM. In particular in our recent work we have shown that in the LHM, electromagnetic (EM) wave propagation is slow with group velocity of 0.02c [9,10] where *c* is the velocity of EM wave in vacuum. This slow group velocity combined with large bandwidth can be used for designing a delay line filter with a large pass band.

In the case of TE mode, for an angle of incidence of 60° the refractive index is found to vary from *0 to -0.48*, which is a 200% change for a frequency change of 42%. Such a large $dn/d\omega$ results in a large $d\phi/d\omega$, which can be used in designing ultra sensitive phase shifters.

In conclusion, negative refraction is demonstrated for both TM and TE mode propagation in a metallic PC. Almost 400% improvement in the bandwidth for negative refraction is obtained for TE mode propagation in this tailor-made PC. The strong dispersion observed results in a change of 200% in the negative refractive index for a frequency change of 42%. The ease and low cost of fabrication of metallic PC vis- a-vis a dielectric PC and metamaterials make them ideal for a wide range of applications. Precise control over the geometry, choice of mode and scalability to submicrometer dimensions of PCs shows promise for applications from microwave to optical frequencies.

We thank John Derov and Beverly Turchinetz for invaluable contributions. The Air Force Research Laboratories, Hanscom AFB and the National Science Foundation supported this work under contract No. F33615-01-1-1007.

[1] R. A. Shelby, D. R. Smith, and S. Schultz, Science **292**, 77 (2001).
[2] M. Notomi, Phys. Rev. B **62**, 10696 (2002); Opt. Quantum Electron. **34**, 133 (2002).
[3] C. Luo, S. G. Johnson, and J. D. Joannopoulos, Phys. Rev. B **65**, 201104 (2002).
[4] J. B. Pendry, Phys. Rev. Lett. **85**, 3966 (2000).
[5] P. V. Parimi, W. T. Lu, P. Vodo, and S. Sridhar, Nature(London) **426**, 404 (2003).
[6] P. V. Parimi, W. T. Lu, P. Vodo, J. B. Sokoloff, and S. Sridhar, Phys. Rev. Lett. 92, 127401-1 (2004).
[7] E. Cubukcu, K. Aydin, E. Ozbay, S. Foteinopoulou, and C. M. Soukoulis, Nature (London) **423**, 604 (2003).
[8] C. G. Parazzoli, R. B. Greegor, K. Li, B. E. C. Koltenbach, and M. C. Tanielian, Phys. Rev. Lett. **90**, 107401 (2003).
[9] P. V. Parimi, P. Vodo, W. T. Lu, J. S. Derov, B. Turchinetz, and S. Sridhar, unpublished
[10] E. DiGennaro, P. V. Parimi, W. T. Lu, and S. Sridhar, unpublished